# Embedding Population Dynamics Models in Inference

**Stephen T. Buckland, Ken B. Newman, Carmen Fernández, Len Thomas and John Harwood**

*Abstract.* Increasing pressures on the environment are generating an ever-increasing need to manage animal and plant populations sustainably, and to protect and rebuild endangered populations. Effective management requires reliable mathematical models, so that the effects of management action can be predicted, and the uncertainty in these predictions quantified. These models must be able to predict the response of populations to anthropogenic change, while handling the major sources of uncertainty. We describe a simple "building block" approach to formulating discrete-time models. We show how to estimate the parameters of such models from time series of data, and how to quantify uncertainty in those estimates and in numbers of individuals of different types in populations, using computer-intensive Bayesian methods. We also discuss advantages and pitfalls of the approach, and give an example using the British grey seal population.

## 1. INTRODUCTION

At the 2002 World Summit on Sustainable Development in Johannesburg, political leaders agreed to strive for "a significant reduction in the current rate of loss of biological diversity" by the year 2010. Initial steps to achieve this include developing and implementing monitoring programs and data collection procedures that quantify the rate of loss of biodiversity (Buckland, Magurran, Green and Fewster, **?**yearBMGF05), allowing assessment of the success of management actions. However, monitoring is a blunt instrument for management, because of the long lag between action and observed effects. Explanatory mathematical models provide "what if" tools for *predicting* the impact of different management actions on populations and biodiversity *before* a course of action is chosen. Both elements are needed: models, to guide and increase the effectiveness of management action; and monitoring, to provide a retrospective measure of whether the predicted effects have been achieved. Indeed, the two approaches should be fully integrated, by using the data from monitoring programs to update and fine-tune the mathematical models. One way to integrate the two elements is through the use of adaptive management techniques (Walters, 2002).

*Stephen T. Buckland is Professor of Statistics, Centre for Research into Ecological and Environmental Modelling, University of St Andrews, St Andrews KY16 9LZ, Scotland e-mail: steve@mcs.st-and.ac.uk. Ken B. Newman is Mathematical Statistician, U.S. Fish and Wildlife Service, Stockton, California 95205, USA e-mail: Ken_Newman@fws.gov. Carmen Fernández is Research Scientist, Instituto Español de Oceanografía, Cabo Estai–Canido, Apdo. 1552, 36200 Vigo, Spain e-mail: carmen.fernandez@vi.ieo.es. Len Thomas is RCUK Academic Fellow, Centre for Research into Ecological and Environmental Modelling, University of St Andrews, St Andrews KY16 9LZ, Scotland e-mail: len@mcs.st-and.ac.uk. John Harwood is Director, Centre for Research into Ecological and Environmental Modelling, University of St Andrews, St Andrews KY16 9LZ, Scotland e-mail: jh17@st-and.ac.uk.*

The process of creating explanatory mathematical models involves several steps, including formulating one or more models, fitting the models to data, and, in situations of multiple models, selecting or averaging models that will be used for prediction (Burnham and Anderson, 2002). Realistic models that meet the needs of policymakers and biodiversity managers often have a high degree of complexity. If these people are not to be misled, it is essential that uncertainty in these models is quantified (Harwood and Stokes, 2003; Clark and Bjørnstad, 2004). The uncertainty in modeling population dynamics that arises includes four main sources: process variation (demographic and environmental stochasticity), observation error, parameter uncertainty and model uncertainty.

A popular and powerful approach to formulating models for plant and animal population dynamics has been the use of matrix models (Caswell, 2001). Such models explicitly account for the various processes that affect the dynamics of populations, for example, survival, birth, movement. However, fitting such matrix models to data, for example, estimating survival and fecundity rates, has often been carried out in a somewhat piecemeal fashion, and uncertainty about model parameters has not always been incorporated into model projections, let alone uncertainty associated with the initial choice of the particular matrix model.

In contrast, many statistical models for time series of population data can readily be fit in an integrated manner and various forms of uncertainty simultaneously accounted for. However, model formulation is often empirical, with no attempt to incorporate the processes underlying population dynamics explicitly, for example, AR($p$) and ARIMA models.

In this paper we review recent developments in modeling population dynamics and describe an integrated approach to formulating, fitting and selecting realistic models for population dynamics that builds upon the matrix model framework. The resulting models are not necessarily matrix models, and are in fact more flexible. The approach is embedded within a Bayesian inferential framework, so that the main sources of uncertainty are accommodated. Two model fitting approaches are discussed, Markov chain Monte Carlo (MCMC) and sequential importance sampling (SIS), and an example of the approach is given for the British grey seal population.

## 2. RECENT DEVELOPMENTS IN MODELING POPULATION DYNAMICS

Matrix population models have long been used for describing population dynamics. Following the pioneering work of Leslie (1945, 1948), such models were usually deterministic; if observational data were used at all for fitting models, it was for *ad hoc* estimation of parameters of the population model, to allow deterministic projection of the population of interest. Caswell (2001) gives a comprehensive account of the mathematical development of the topic, which includes stochastic extensions (demographic and environmental variation), and asymptotic analyses of growth rates and age and stage class distributions for deterministic and stochastic matrix models. However, in many cases an integration of field or laboratory experiment data with an underlying population dynamics model is lacking. Quoting Tuljapurkar (1997) following an exposition of analysis of stochastic matrix models: "A useful application of stochastic models that we have not discussed here is the development of methods for estimating the vital rates of structured populations from data. Such estimation methods … are desirable in ecology, although rarely used."

An integration of data with population dynamics models, and an embedding of such models in statistical inference, is needed to allow for simultaneous accounting of uncertainties about model parameter values, observation error, process variation (demographic and environmental stochasticity) and model uncertainty. State-space models (Harvey, 1989; West and Harrison, 1997), particularly when applied in a Bayesian framework, are a means of integrating data with population dynamics models and readily quantifying the various types of uncertainty (Calder, Lavine, Müller and Clark, 2003; Clark, Ferraz, Oguge, Hays and DiCostanzo, 2005). A process model specifies probability distributions associated with the relevant population processes, such as birth, survival and movement, and a corresponding observation model defines the probability distribution of observations, relating them to states. Using the terminology of Caswell (2001, pages 37–38), individuals are classified according to $i$-states (e.g., age category, sex, species). The vector of counts (or biomass) of individuals by $i$-state is the $p$-state of the population, that is, the distribution of $i$-states in the population. The elements of the state vector in the process model may correspond to individuals or to groups of individuals. State-space models are first-order Markov: the state at time $t$ depends on the state at time $t-1$, but is conditionally independent of earlier states. Higher-order Markov models are readily accommodated, and Newman, Buckland, Lindley, Thomas and Fernández (2006) use the term "hidden process models" to describe the entire range of such models.

Lavine, Beckage and Clark (2002) provide an example of a complex process model for tree seedling mortality which could be formulated as a state equation. The observation equation in their case is degenerate in the sense that there is (assumed to be) no observation error: the true numbers of seedlings are recorded on the plots. Thus the observation equation would equate observations to the sum of counts for indistinguishable types, while the state equation would split out these types—hence it is an example of a hidden process model. If inference were to be drawn on a wider population beyond the sampled plots, then the observation equation would not be degenerate.

Many of the key developments in this field have been made in the context of fisheries stock assessment; Quinn and Deriso (1999) give a detailed account of these developments. Early methods typically modeled either process variation or (more usually) observation error (Hilborn and Walters, 1992, page 226). An exception was Collie and Sissenwine (1983), who modeled a time series of relative abundance data with observation error to estimate population size. Process variation was incorporated by assuming that survival rate was a random variable. Mendelssohn (1988) extended their approach to allow the underlying population dynamics to be random. He obtained maximum likelihood estimates of the model parameters using the Kalman filter (Kalman, 1960) together with the EM algorithm (Dempster, Laird and Rubin, 1977). Gudmundsson (1987, 1994), Sullivan (1992) and Schnute (1994) all used the Kalman filter to fit state-space models of fish dynamics to time series of catch data, and Newman (1998) incorporated a spatial component. All of these authors adopted linear models (or linear approximations to nonlinear models), and assumed that process variation and observation errors were both normally distributed.

Nonlinear, nonnormal models have been developed within a Bayesian framework. Hilborn, Pikitch and McAllister (1994), McAllister, Pikitch, Punt and Hilborn (1994) and Schnute (1994) developed Bayesian approaches in a fisheries context, and McAllister and Ianelli (1997) compared fitting algorithms based on sequential importance sampling (SIS, also known as particle filtering), Markov chain Monte Carlo (MCMC) and adaptive importance sampling. Millar and Meyer (2000) developed an MCMC fitting algorithm, which Meyer and Millar (1999) implemented in the BUGS package, thus making the methods more accessible to the user community. Rivot, Prévost, Parent and Baglinière (2004) also used BUGS to fit their model. Cunningham, Reid, McAllister, Kirkwood and Darby (2007) used the sampling importance resampling (SIR) algorithm (Rubin, 1988) to model three mackerel stocks, spread through seven geographic regions, with deterministic movement between them.

Outside of fisheries, Raftery, Givens and Zeh (1995) used a "Bayesian synthesis" approach to draw inference from a deterministic population dynamics model for bowhead whales. As pointed out by Wolpert (1995), their approach suffers from Borel's paradox: their results are dependent on the (nonunique) parameterization chosen for the population dynamics model. Subsequently, Poole and Raftery (2000) developed a "Bayesian melding" method, which is coherent. Trenkel, Elston and Buckland (2000) developed a model of red deer population dynamics that managers could use to explore the likely effects of different culling strategies. They used SIS with kernel smoothing to fit the model in a Bayesian framework. Besbeas, Freeman, Morgan and Catchpole (2002), Besbeas, Lebreton and Morgan (2003) and Besbeas, Freeman and Morgan (2005) used the Kalman filter to fit state-space models to a combination of abundance and demographic data on two species of birds (grey heron and northern lapwing). MCMC has also been used to fit state-space models for bird populations (Wikle, 2003; Clark and Bjørnstad, 2004) and moose (Clark and Bjørnstad, 2004). Thomas, Buckland, Newman and Harwood (2005) used SIS to model the dynamics of a spatially structured grey seal population, allowing estimation of movement rates between regions. Lele (2006) used the composite-likelihood method for estimating the parameters of the Gompertz model in the presence of sampling variability.

To date, model uncertainty has received little attention. Bayesian model averaging may be incorpo-rated into importance sampling as noted by Buckland, Newman, Thomas and Koesters (2004) and Cunningham et al. (2007). Several models are defined, and a prior is specified reflecting the initial belief on the relative plausibility of each model. In the absence of better knowledge, we would assign each model equal prior weight. If MCMC is used, then reversible jump MCMC (Green, 1995) can be used to sample what may be very high-dimensional model space using the approach of King and Brooks (2002a, b).

## 3. MODELING POPULATION PROCESSES AND ASSOCIATED INFERENCE

### 3.1 Leslie and Lefkovitch Matrices Decomposed

A Leslie matrix (Leslie, 1945, 1948; Caswell, 2001) is a population projection matrix that shows how, in the absence of process variation, a population updates itself from one year to the next, through births, deaths and aging. We use the term "generalized Leslie matrix" to indicate an age-structured population projection matrix with at least one additional process, such as movement or sex assignment. If, instead of aging, we have stages of development, then the population projection matrix is usually referred to as a Lefkovitch matrix (Lefkovitch, 1965; Caswell, 2001). In the following, we show how such matrices can be expressed as products of simpler matrices, each one corresponding to a single biological process (Buckland et al., 2004). This approach has been used in the context of deterministic matrix population models by Lebreton (1973) and Lebreton and Isenmann (1976). Hooten, Wikle, Dorazio and Royle (2007) used the same strategy to split their matrix model for the growth and spread of the US population of Eurasian collared-doves into a growth process and a movement process.

Suppose that we have one species of animal, divided into two age classes, with a total of $n_{0,t-1}$ newly born individuals and $n_{1,t-1}$ adults in the population at the end of year $t-1$. Suppose further that the population is subject to just three processes: survival through the next year, age incrementation and births. For simplicity, we model the females in the population only. The expected number of survivors to the end of the next year can then be written

$$\begin{pmatrix} E(s_{0,t}) \\ E(s_{1,t}) \end{pmatrix} = \mathbf{S} \begin{pmatrix} n_{0,t-1} \\ n_{1,t-1} \end{pmatrix} = \begin{pmatrix} \phi_0 & 0 \\ 0 & \phi_1 \end{pmatrix} \begin{pmatrix} n_{0,t-1} \\ n_{1,t-1} \end{pmatrix},$$

where $\phi_0$ is the probability of survival of newly born individuals to the end of their first year, $\phi_1$ is the annual survival probability of adults and $\mathbf{S}$ is a survival projection matrix. We can introduce demographic stochasticity by specifying

$$\begin{pmatrix} s_{0,t} \sim \text{binomial}(n_{0,t-1}, \phi_0) \\ s_{1,t} \sim \text{binomial}(n_{1,t-1}, \phi_1) \end{pmatrix}.$$

We could also introduce environmental stochasticity by allowing the survival probabilities $\phi_0$ and $\phi_1$ to vary at random between years. Survival probabilities could also be modeled as functions of covariates, for example, using logistic functions, as in Buckland et al. (2004). If some of these covariates relate to the individual, such as body mass or condition, then $i$-states (the states of individuals in the population) can be modeled. If relevant covariates are unavailable, but survival probabilities are variable, they could be modeled using random effects.

Age incrementation is deterministic, so that the numbers of individuals immediately before births take place are

$$\begin{pmatrix} 0 \\ a_{1,t} \end{pmatrix} = \mathbf{A} \begin{pmatrix} s_{0,t} \\ s_{1,t} \end{pmatrix} = \begin{pmatrix} 0 & 0 \\ 1 & 1 \end{pmatrix} \begin{pmatrix} s_{0,t} \\ s_{1,t} \end{pmatrix},$$

where $\mathbf{A}$ is an aging projection matrix. Finally, we can specify the birth process by writing

$$\begin{pmatrix} E(n_{0,t}) \\ n_{1,t} \end{pmatrix} = \mathbf{B} \begin{pmatrix} 0 \\ a_{1,t} \end{pmatrix} = \begin{pmatrix} 1 & \lambda \\ 0 & 1 \end{pmatrix} \begin{pmatrix} 0 \\ a_{1,t} \end{pmatrix},$$

where $\mathbf{B}$ is a birth projection matrix, with a suitable stochastic model for births, for example, $n_{0,t} \sim \text{Poisson}(\lambda a_{1,t})$. (Note that the first element of the birth matrix could equivalently be set to zero, since after aging and before new births occur, there are no young in the population; we set it to unity in anticipation of the growth model below.)

Thus, conditional on numbers at the end of year $t-1$, the expected numbers of individuals present at the end of year $t$ are

$$\begin{pmatrix} E(n_{0,t}) \\ E(n_{1,t}) \end{pmatrix} = \begin{pmatrix} 1 & \lambda \\ 0 & 1 \end{pmatrix} \begin{pmatrix} 0 & 0 \\ 1 & 1 \end{pmatrix} \begin{pmatrix} \phi_0 & 0 \\ 0 & \phi_1 \end{pmatrix} \begin{pmatrix} n_{0,t-1} \\ n_{1,t-1} \end{pmatrix}$$
$$= \begin{pmatrix} \lambda\phi_0 & \lambda\phi_1 \\ \phi_0 & \phi_1 \end{pmatrix} \begin{pmatrix} n_{0,t-1} \\ n_{1,t-1} \end{pmatrix}.$$

We denote this Model 1. The vector $\mathbf{n}_t = \binom{n_{0,t}}{n_{1,t}}$ is termed the $p$-state or simply the state vector, because it tallies the number of individuals in each state (mature or immature here). The product of matrices is an example of a simple Leslie matrix.

Note, however, that only survivors can breed; in the traditional Leslie matrix, the first row would not contain the survival parameters, so that the birth rate applies to all individuals, whether or not they survive until the breeding season. By varying the sequencing of the processes that alter the population, one can readily construct variations on classic Leslie matrix formulations. Note that the above equation may be expressed in matrix form as $E(\mathbf{n}_t|\mathbf{n}_{t-1}) = \mathbf{P}\mathbf{n}_{t-1}$, with the Leslie population projection matrix $\mathbf{P} = \mathbf{BAS} = \begin{pmatrix} \lambda\phi_0 & \lambda\phi_1 \\ \phi_0 & \phi_1 \end{pmatrix}$.

Suppose we wish to model two growth stages rather than two age classes, and the probability that an individual moves from stage 1 to stage 2 is $\pi$. Then the age incrementation matrix $\mathbf{A} = \begin{pmatrix} 0 & 0 \\ 1 & 1 \end{pmatrix}$ is replaced by the growth matrix $\mathbf{G} = \begin{pmatrix} 1-\pi & 0 \\ \pi & 1 \end{pmatrix}$ and the product of the three matrices is now a Lefkovitch population projection matrix: $\mathbf{P} = \mathbf{BGS} = \begin{pmatrix} (1-\pi+\lambda\pi)\phi_0 & \lambda\phi_1 \\ \pi\phi_0 & \phi_1 \end{pmatrix}$. Denote this by Model 2.

We can now extend Model 1 to have two subpopulations with movement rate $\mu$ between them; we label this Model 3. If movement occurs just before the breeding season, then

$$\begin{pmatrix} E(n_{01,t}) \\ E(n_{11,t}) \\ E(n_{02,t}) \\ E(n_{12,t}) \end{pmatrix} = \begin{pmatrix} 0 & \lambda & 0 & 0 \\ 0 & 1 & 0 & 0 \\ 0 & 0 & 0 & \lambda \\ 0 & 0 & 0 & 1 \end{pmatrix} \begin{pmatrix} 0 & 0 & 0 & 0 \\ 1 & 1 & 0 & 0 \\ 0 & 0 & 0 & 0 \\ 0 & 0 & 1 & 1 \end{pmatrix}$$
$$\cdot \begin{pmatrix} 1-\mu & 0 & \mu & 0 \\ 0 & 1-\mu & 0 & \mu \\ \mu & 0 & 1-\mu & 0 \\ 0 & \mu & 0 & 1-\mu \end{pmatrix}$$
$$\cdot \begin{pmatrix} \phi_0 & 0 & 0 & 0 \\ 0 & \phi_1 & 0 & 0 \\ 0 & 0 & \phi_0 & 0 \\ 0 & 0 & 0 & \phi_1 \end{pmatrix} \begin{pmatrix} n_{01,t-1} \\ n_{11,t-1} \\ n_{02,t-1} \\ n_{12,t-1} \end{pmatrix},$$

where $n_{0j,t}$ indicates number of immature individuals in subpopulation $j$ in year $t$, $j = 1, 2$, and similarly for adults. This may be expressed in matrix form as $E(\mathbf{n}_t|\mathbf{n}_{t-1}) = \mathbf{BAMSn}_{t-1}$, where the birth ($\mathbf{B}$), aging ($\mathbf{A}$), movement ($\mathbf{M}$) and survival ($\mathbf{S}$) projection matrices are as above. The product of matrices is now a generalized Leslie matrix:

$$\mathbf{P} = \mathbf{BAMS}$$
$$= \begin{pmatrix} \lambda\phi_0(1-\mu) & \lambda\phi_1(1-\mu) & \lambda\phi_0\mu & \lambda\phi_1\mu \\ \phi_0(1-\mu) & \phi_1(1-\mu) & \phi_0\mu & \phi_1\mu \\ \lambda\phi_0\mu & \lambda\phi_1\mu & \lambda\phi_0(1-\mu) & \lambda\phi_1(1-\mu) \\ \phi_0\mu & \phi_1\mu & \phi_0(1-\mu) & \phi_1(1-\mu) \end{pmatrix}.$$

Diagrammatic representations of Models 1–3 are given in the Appendix.

Caswell (2001, page 110) considers an additive decomposition of the projection matrix, in which one of the matrices describes reproduction and the other transitions. The matrix for reproduction comprises elements that represent the expected number of offspring to individuals, where columns represent the type of individual, and rows the type of offspring. The transition matrix comprises elements that represent the probability that an individual of type $j$ (represented by columns) at time $t-1$ is in the population and of type $i$ (represented by rows) at time $t$. For the above multiplicative decompositions, each process can be specified by its own matrix, whereas in Caswell's additive decomposition, each process must be included as a component of either reproduction or transition. Thus for a model with growth stages and two sexes, survival and growth processes both appear in the transition matrix, and birth and sex assignment are both handled in the reproduction matrix. The multiplicative decomposition therefore offers an easier formulation of the model. It also explicitly orders processes in time, whereas the additive decomposition does not. The latter approach therefore shares a disadvantage with the standard Leslie matrix: the breeding rate applies to all individuals, whether or not they survive to the breeding season. (If the year is redefined to start just before the breeding season, then first-year survival cannot be modeled using an additive decomposition.) A further disadvantage of the additive decomposition is that it does not allow the number of states within the population to vary through the year, whereas this is readily modeled in the sequential multiplicative framework (Buckland et al., 2004), allowing greater flexibility.

## 3.2 The State Equation

Denote the state vector after process $k$ has occurred in year $t$ by $\mathbf{u}_{k,t}$, where $k = 1, \dots, K$, say, and $t = 1, \dots, T$. Then $\mathbf{u}_{K,t} \equiv \mathbf{n}_t$. For Model 1, for example, $K = 3$; $\mathbf{u}_{1,t}$ represents numbers of survivors in each $i$-state through to the end of year $t$, $\mathbf{u}_{2,t}$ represents numbers of adults after age incrementation, and $\mathbf{u}_{3,t} \equiv \mathbf{n}_t$. The state equation for process $k$ in year $t$ specifies how the states are updated. Assuming the process is first-order Markov (an assumption that may easily be relaxed), $\mathbf{u}_{k,t} = \tilde{\mathbf{P}}_{k,t}(\mathbf{u}_{k-1,t})$ for $k = 1, \dots, K$, where $\mathbf{u}_{0,t} \equiv \mathbf{n}_{t-1} \equiv \mathbf{u}_{K,t-1}$, and $\tilde{\mathbf{P}}_{k,t}(\cdot)$ is an operator to be specified, which is random if the process is stochastic. For the general case, the state equation may be expressed as

$$\mathbf{n}_t = \tilde{\mathbf{P}}_t(\mathbf{n}_{t-1}), \quad (1)$$

where $\tilde{\mathbf{P}}_t(\cdot)$ is the composition

$$\tilde{\mathbf{P}}_t(\cdot) = \tilde{\mathbf{P}}_{K,t}(\tilde{\mathbf{P}}_{K-1,t}(\cdots \tilde{\mathbf{P}}_{1,t}(\cdot)\cdots)).$$

Matrix population models in which all processes are deterministic arise as special cases. Then we can write

$$\mathbf{n}_t = \mathbf{P}_t \mathbf{n}_{t-1},$$

where $\mathbf{P}_t = \mathbf{P}_{K,t}\mathbf{P}_{K-1,t}\cdots\mathbf{P}_{1,t}$ is typically a generalized Leslie or Lefkovitch matrix. Note that $\mathbf{P}_t$ may be a function of $\mathbf{n}_{t-1}$, allowing models that are nonlinear in the states, so that, for example, rates can be a function of population size (Caswell, 2001). A second special case is when at least one process is stochastic, but the expected values of the elements of $\mathbf{n}_t$ may be expressed as functions of the elements of $\mathbf{n}_{t-1}$. Then

$$E(\mathbf{n}_t|\mathbf{n}_{t-1}) = \mathbf{P}_t \mathbf{n}_{t-1}, \tag{2}$$

where $\mathbf{P}_t$ is a population projection matrix such that $E(\tilde{\mathbf{P}}_t(\mathbf{n}_{t-1})|\mathbf{n}_{t-1}) = \mathbf{P}_t\mathbf{n}_{t-1}$. Examples are given by Models 1–3. For Model 1, $\tilde{\mathbf{P}}_{1,t}(\cdot)$ is a random variable generated from a binomial distribution corresponding to each $i$-state, $\tilde{\mathbf{P}}_{2,t}(\cdot)$ is deterministic, and $\tilde{\mathbf{P}}_{3,t}(\cdot)$ has a stochastic Poisson-distributed component and a deterministic one. In this case, $E[\tilde{\mathbf{P}}_{3,t}(\tilde{\mathbf{P}}_{2,t}(\tilde{\mathbf{P}}_{1,t}(\mathbf{n}_{t-1})))|\mathbf{n}_{t-1}] = \mathbf{BASn}_{t-1}$ so that $\mathbf{P}_t = \mathbf{BAS}$. We stress that the model fitting procedures described here do not require (2) to hold and are applicable in the wider context of (1).

For a full specification of the process model, we need to choose a probability density function (p.d.f.) for the initial state vector, say $g_0(\mathbf{n}_0|\boldsymbol{\theta})$, and a p.d.f. for the state vector at time $t$ given the state vector at time $t-1$, $g_t(\mathbf{n}_t|\mathbf{n}_{t-1},\boldsymbol{\theta})$, where $\boldsymbol{\theta}$ is the vector of unknown parameters. This p.d.f. is determined by the $K$ p.d.f.'s corresponding to each of the processes in (1) (Buckland et al., 2004). We note that, in the case of multiple processes, the resulting p.d.f. is a composite function that can also include convolutions. Direct evaluation of the p.d.f. can thus be quite difficult without the inclusion of intermediate latent states (Clark et al., 2005; Newman, Fernández, Buckland and Thomas, 2007); Lele (2006) addresses a similar problem within a likelihood framework.

## 3.3 The Observation Equation

The observation equation relates the observation vector $\mathbf{y}_t$ to the states at time $t$ through an operator $\tilde{\mathbf{O}}_t(\cdot)$, so that $\mathbf{y}_t = \tilde{\mathbf{O}}_t(\mathbf{n}_t)$. We denote the corresponding observation p.d.f. by $f_t(\mathbf{y}_t|\mathbf{n}_t,\boldsymbol{\theta})$. If the operator is random but linear, then $E(\mathbf{y}_t|\mathbf{n}_t) = \mathbf{O}_t\mathbf{n}_t$ for the appropriate matrix $\mathbf{O}_t$. For example, if for Model 3 we have an estimate $y_{j,t}$ of total population size just after breeding for subpopulation $j$, with corresponding variance $\sigma^2_{j,t}$, $j = 1, 2$, and the two estimates are independently normally distributed, then

$$E(\mathbf{y}_t|\mathbf{n}_t) = \begin{pmatrix} E(y_{1,t}) \\ E(y_{2,t}) \end{pmatrix} = \begin{pmatrix} 1\,1\,0\,0 \\ 0\,0\,1\,1 \end{pmatrix} \begin{pmatrix} n_{01,t} \\ n_{11,t} \\ n_{02,t} \\ n_{12,t} \end{pmatrix}$$

and

$$f_t(\mathbf{y}_t|\mathbf{n}_t,\boldsymbol{\theta}) = \prod_{j=1}^{2} \frac{1}{\sqrt{2\pi\sigma^2_{j,t}}} \exp\left[\frac{\{y_{j,t} - E(y_{j,t})\}^2}{2\sigma^2_{j,t}}\right].$$

In a Bayesian context, a prior distribution for $\sigma^2_{j,t}$ may be based on an estimate $\hat{\sigma}^2_{j,t}$, together with its estimated precision.

Unknown parameters may feature in the projection matrix $\mathbf{O}_t$. In the above example, if the observations were incomplete counts, with probability $p$ that any individual animal was counted, we might have

$$E(\mathbf{y}_t|\mathbf{n}_t) = \begin{pmatrix} E(y_{1,t}) \\ E(y_{2,t}) \end{pmatrix} = \begin{pmatrix} p\,p\,0\,0 \\ 0\,0\,p\,p \end{pmatrix} \begin{pmatrix} n_{01,t} \\ n_{11,t} \\ n_{02,t} \\ n_{12,t} \end{pmatrix}$$

and

$$f_t(\mathbf{y}_t|\mathbf{n}_t,\boldsymbol{\theta}) = \prod_{j=1}^{2} \binom{n_{0j,t} + n_{1j,t}}{y_{j,t}} p^{y_{j,t}} (1-p)^{n_{0j,t}+n_{1j,t}-y_{j,t}}.$$

Given additional independent data for estimating $p$, the observation p.d.f. may be expanded by multiplying by the likelihood corresponding to those data.

## 3.4 Bayesian Inference

Within a Bayesian framework, denote the prior distribution of the parameters $\boldsymbol{\theta}$ by $g(\boldsymbol{\theta})$. Then the joint prior distribution for the state vector $\mathbf{n}_t$ and the parameters $\boldsymbol{\theta}$ is

$$g(\boldsymbol{\theta}) \times g_0(\mathbf{n}_0|\boldsymbol{\theta}) \times \prod_{t=1}^{T} g_t(\mathbf{n}_t|\mathbf{n}_{t-1},\boldsymbol{\theta}),$$

and the likelihood is

$$\prod_{t=1}^{T} f_t(\mathbf{y}_t|\mathbf{n}_t,\boldsymbol{\theta}).$$

Hence the posterior distribution is

$$g(\mathbf{n}_0,\ldots,\mathbf{n}_T,\boldsymbol{\theta}|\mathbf{y}_1,\ldots,\mathbf{y}_T) = \Bigl(g(\boldsymbol{\theta}) \times g_0(\mathbf{n}_0|\boldsymbol{\theta}) \times \prod_{t=1}^{T}\{g_t(\mathbf{n}_t|\mathbf{n}_{t-1},\boldsymbol{\theta}) \times f_t(\mathbf{y}_t|\mathbf{n}_t,\boldsymbol{\theta})\}\Bigr) \times (f(\mathbf{y}_1,\ldots,\mathbf{y}_T))^{-1}. \tag{3}$$

This approach for modeling complex environmental and ecological processes was espoused by Berliner (1996), Wikle, Berliner and Cressie (1998) and Wikle (2003).

At time $t \le T$, the following types of inference are often useful: filtering, using $g(\mathbf{n}_t,\boldsymbol{\theta}|\mathbf{y}_1,\ldots,\mathbf{y}_t)$ (i.e., only observations up to time $t$ are used); smoothing, using $g(\mathbf{n}_t,\boldsymbol{\theta}|\mathbf{y}_1,\ldots,\mathbf{y}_T)$ (i.e., the full time series of observations up to time $T$ is used to estimate the state vector at time $t$); and prediction, using $g(\mathbf{n}_{t'},\boldsymbol{\theta}|\mathbf{y}_1,\ldots,\mathbf{y}_t)$ (i.e., observations up to year $t$ are used to predict the state vector in year $t'$, $t' > t$).

## 3.5 Model Fitting

SIS and MCMC are both computer-intensive Monte Carlo methods, useful for fitting complex models of the type described above, usually within a Bayesian framework. Liu (2001) describes both SIS and MCMC, and Gilks, Richardson and Spiegelhalter (1996) and Doucet, de Freitas and Gordon (2001) include a range of articles on MCMC and SIS, respectively.

Newman et al. (2006, 2007) give full details of how the two approaches may be applied to fit models that fall within the above framework. Because SIS is less widely used than MCMC, we start with a brief description of it.

3.5.1 *Sequential importance sampling.* In its simplest form, a large number $R$ of "particles" are simulated from the prior distribution $g(\mathbf{n}_0,\boldsymbol{\theta}) = g(\boldsymbol{\theta}) \times g_0(\mathbf{n}_0|\boldsymbol{\theta})$. The $r$th "particle" represents a single prior realization $\boldsymbol{\theta}_r$ of the parameters of the population model, together with a single realization of the population in the initial year, $\mathbf{n}_{0,r}$. Each particle is projected forward to time $t = 1$, by simulating the state vector $\mathbf{n}_{1,r}$ from $g_1(\mathbf{n}_1|\mathbf{n}_{0,r},\boldsymbol{\theta}_r)$ for all $r$. Particles are resampled with weights proportional to the contribution to the likelihood of any observations at $t = 1$. Thus the weights are $w_r = f_1(\mathbf{y}_1|\mathbf{n}_{1,r},\boldsymbol{\theta}_r)/\sum_{r=1}^{R} f_1(\mathbf{y}_1|\mathbf{n}_{1,r},\boldsymbol{\theta}_r)$. This resampling, known as bootstrap filtering, was introduced by Gordon, Salmond and Smith (1993). The surviving particles $(\boldsymbol{\theta}_r,\mathbf{n}_{0,r},\mathbf{n}_{1,r})$ are approximately a sample from the posterior distribution, given the data at time $t = 1$. They may be projected forward to $t = 2$, with the state vector simulated from $g_2(\mathbf{n}_2|\mathbf{n}_{1,r},\mathbf{n}_{0,r},\boldsymbol{\theta}_r)$, and so on, until resampling at time $T$ has been carried out. The surviving particles are then an approximate sample from the posterior distribution given by (3).

A practical problem with the above implementation of SIS is "particle depletion": typically, only a tiny proportion of original particles survives, and most of those appear in the posterior sample many times. Consequently, Monte Carlo variation in the sample posterior distribution between different simulations can be quite high for this simplest form of sequential importance sampling. The problem is more acute for the static parameters $\boldsymbol{\theta}$ and the states corresponding to the early years. Various tricks are used to mitigate the effects of particle depletion, such as sampling from a proposal distribution that more closely matches the posterior than does the prior (Liu and Chen, 1998), kernel smoothing (West, 1993a, 1993b; Trenkel, Elston and Buckland, 2000), the auxiliary particle filter (Pitt and Shephard, 1999; Thomas et al., 2005), residual sampling and partial rejection control (Liu, 2001).

3.5.2 *Markov chain Monte Carlo.* This method generates dependent samples from the posterior distribution in (3), by simulating a Markov chain whose stationary distribution is the required posterior distribution. A particular strength of this method is that it allows the typically large dimensionality of the posterior distribution to be broken down, by simulating blocks of variables (parameters or states) in turn, conditioning on the current values of the other variables in the chain. An efficient implementation of MCMC for fitting population dynamics models depends critically on the choice of blocking scheme for the parameters and states, and of proposal distributions (Gilks and Roberts, 1996). Reparameterization can be very useful in some cases.

## 3.6 Extending the Model Framework

The modeling framework captured by (1) is very flexible, and goes far beyond the simple examples presented in Section 3.1. We now briefly discuss some of the possibilities.

For moderate and large populations, demographic stochasticity may be negligible compared to environmental stochasticity. Vital rates (i.e., birth and survival rates) can be modeled as (autocorrelated) random variables to account for environmental stochasticity or trends in habitat quality (Rivot et al., 2004). Vital rates could also vary in a completely random manner (i.i.d. sequences; Newman, 2000), evolve according to a discrete-time Markov chain, or follow an autoregressive-moving-average process (Caswell, 2001, pages 378–379; see also Johnson and Hoeting, 2003). Further, hierarchical models can be used that include random effects for sampling correlated processes (Clark et al., 2005). More detailed models could specify vital rates as functions of covariates, such as the environment, habitat or age, which could vary stochastically. These covariates could also reflect deliberate management actions, or accidental anthropogenic changes in the environment, allowing their effects on population abundance to be predicted. Inclusion of covariates that relate to $i$-state allows modeling of individuals. Modeling vital rates as random effects is another way to allow for individual variation. Density-dependent effects (i.e., effects that reduce birth and/or survival rates as the population gets larger) can be introduced by making vital rates a function of population size. Harvesting and assignment of sex and genotype to newborn animals can also be incorporated.

At a more ambitious level, environmental and demographic processes can be modeled together, rather than simply using environmental variables as covariates in models for population-level parameters. For example, Wikle, Berliner and Cressie (1998) developed a hierarchical space–time model for monthly temperatures and Wikle (2003) developed a hierarchical space–time model for the spread of the house finch. In principle, assuming that temperatures affect house finch survival and subsequent spread, a combined hierarchical space–time model for monthly temperatures and house finch abundance is possible.

Spatial structure can also be modeled, by extending the simple movement model represented by Model 3; the movement rate may also be modeled as a function of animal density and/or covariates (Buckland et al., 2004; Thomas et al., 2005). Similarly, we can include additional species in the model by extending the state vector to include entries corresponding to all species. Survival probabilities and/or birth rates of one species can then be modeled as functions of the abundance of the other species, as in discrete-time predator-prey, consumer-resource or community models (Gurney and Nisbet, 1998).

Many of these possibilities can be viewed as extensions to standard matrix models, or to stochastic linear models as in (2). Table 1 contains some examples, all of which are "hidden" process models because we do not directly observe the states that arise from each process (e.g., numbers of deaths and survivors as a result of the survival process). Instead, we observe or estimate certain states at certain points in time (e.g., number of females present

TABLE 1

*Representations of hidden process models as matrix population models*

| | |
|---|---|
| Randomly varying vital rates: | $E(\mathbf{n}_t \mid \mathbf{n}_{t-1}) = \mathbf{P}_t^{*}\mathbf{n}_{t-1}$ |
| Vital rates as functions of covariates: | $E(\mathbf{n}_t \mid \mathbf{n}_{t-1}) = \mathbf{P}(\mathbf{x}_t)\mathbf{n}_{t-1}$ |
| Density dependence: | $E(\mathbf{n}_t \mid \mathbf{n}_{t-1}) = \mathbf{P}(\mathbf{n}_{t-1})\mathbf{n}_{t-1}$ |
| Metapopulations: | $E\left[\begin{pmatrix}\mathbf{n}_{a,t}\\ \mathbf{n}_{b,t}\\ \vdots\end{pmatrix} \middle| \begin{pmatrix}\mathbf{n}_{a,t-1}\\ \mathbf{n}_{b,t-1}\\ \vdots\end{pmatrix}\right] = \begin{pmatrix}\mathbf{P}_{11} & \mathbf{P}_{12} & \cdots\\ \mathbf{P}_{21} & \mathbf{P}_{22} & \cdots\\ \vdots & \vdots & \ddots\end{pmatrix}\begin{pmatrix}\mathbf{n}_{a,t-1}\\ \mathbf{n}_{b,t-1}\\ \vdots\end{pmatrix}$, |
| where the off-diagonal submatrices handle movement between subpopulations. | |
| Multiple species: | $E(\mathbf{n}_t \mid \mathbf{n}_{t-1}) = E\left[\begin{pmatrix}\mathbf{n}_{a,t}\\ \mathbf{n}_{b,t}\\ \vdots\end{pmatrix} \middle| \begin{pmatrix}\mathbf{n}_{a,t-1}\\ \mathbf{n}_{b,t-1}\\ \vdots\end{pmatrix}\right] = \begin{pmatrix}\mathbf{P}_{11}(\mathbf{n}_{t-1}) & \mathbf{0} & \cdots\\ \mathbf{0} & \mathbf{P}_{22}(\mathbf{n}_{t-1}) & \cdots\\ \vdots & \vdots & \ddots\end{pmatrix}\begin{pmatrix}\mathbf{n}_{a,t-1}\\ \mathbf{n}_{b,t-1}\\ \vdots\end{pmatrix}$ |

*The matrix $\mathbf{P}_t$ is the population projection matrix, showing how expectations of states $\mathbf{n}_t$ conditional on $\mathbf{n}_{t-1}$ relate to $\mathbf{n}_{t-1}$. The matrix $\mathbf{P}_t$ might have elements that are expectations of functions of the vital rates (indicated by $\mathbf{P}_t^{*}$), or might depend on covariates $\mathbf{x}_t$ [denoted $\mathbf{P}(\mathbf{x}_t)$] or states $\mathbf{n}_{t-1}$ [denoted $\mathbf{P}(\mathbf{n}_{t-1})$]. More complex models can be obtained by combining the different model types. Further, $\mathbf{P}_t$ might depend on intermediate states, corresponding to a time point between $t-1$ and $t$, in which case the expectations do not hold and the models should be interpreted in the more general context of (1).

at breeding colonies). Inferences about all the underlying hidden processes, such as plausible parametric forms for models of the processes, and estimates of the corresponding parameters, can be drawn from these observations (Harwood and Stokes, 2003; Thomas et al., 2005; Newman et al., 2006).

As Clark and Bjørnstad (2004) note, missing observations, for example as the result of uneven sampling intervals, are readily accommodated by the state-space framework. Because we split the annual population processes into chronological order, our framework is readily extended to allow observations at different times of the year. Changing effort, another issue considered by Clark and Bjørnstad (2004), is also easily accommodated in this framework: capture or detection probability can be modeled in the observation equation as a function of effort, or counts can be adjusted for effort, so that the adjusted counts (which may be absolute or relative abundance estimates) are entered into the observation equation, along with their estimated precision. Dupuis (1995) and Clark et al. (2005) modeled structured populations in the context of mark-recapture, and Clark et al. (2005) noted that other data models could be used with their methods. In our framework, as with Clark et al. (2005), the structure appears in the state vector, and transitions are modeled in the state equation. For mark-recapture, if the mark status of an animal is included in the state vector, then the capture model is incorporated into the state equation (Buckland et al., 2004), and the observation equation is degenerate: the numbers of animals with capture histories ending with capture are known without error. If, on the other hand, the capture process is considered to be part of the observation process (in which case inference is not conditional on known numbers of individuals captured), it is modeled in the observation equation. Population processes such as survival, births and transitions (e.g., movement) are handled in the state equation, allowing a population dynamics model to be "embedded" in the mark-recapture analysis.

### 3.7 Limitations

The models described here operate in discrete time and space, and the population is partitioned into discrete states. Movement models that are continuous in space can be incorporated, by specifying a probability density function for distance and direction moved, given an animal's current state, location and other relevant covariates. Extension to continuous time is less straightforward. However, we can readily decrease the time intervals that are modeled to reduce the approximations implicit in discretization. Further, for time intervals in which only a single process (e.g., survival) operates, no approximation is required—in effect, we integrate out time over the period. For example, if survival and movement processes operate synchronously, we can specify models for the instantaneous death and movement rates, and use simulation-based methods to embed these continuous-time models in the above structure. However, the benefits of this approach (in terms of improved correspondence with reality) would have to be weighed against the additional computational burden.

The same approach can be used to model communities of species, but our ability to fit such models is hampered by a lack of knowledge and of data on species interactions. For example, one species may act as both predator and prey of another at different stages of their life cycles. Similarly, knowledge is often lacking about the form of the relationships between biological processes and environmental covariates. However, technological advances in data logging offer greater potential for gathering data that can be used to infer such relationships. As computer power increases, we can, for example, contemplate the development of community models that incorporate the important processes, while accommodating the major sources of uncertainty.

## 4. EXAMPLE: MODELING A METAPOPULATION

Thomas et al. (2005) developed a hidden process model of the metapopulation dynamics of British grey seals. These animals spend over 80% of their time at sea (McConnell, Fedak, Lovell and Hammond, 1999), and 90% of this time underwater (Thompson, Hammond, Nicholas and Fedak, 1991), so that it is difficult to survey the entire population. However, the species breeds colonially and pups spend most of the first three weeks of their lives ashore, where they are readily counted. Aerial pup counts have been conducted at all of the major Scottish breeding colonies in every year since 1984. These counts are used to estimate the total number of pups born at each colony in each year.

To illustrate the ideas of this paper, we modeled the population of female seals for the period 1984–2002 by aggregating the colonies into four geographically distinct regions: North Sea (4 colonies) Inner

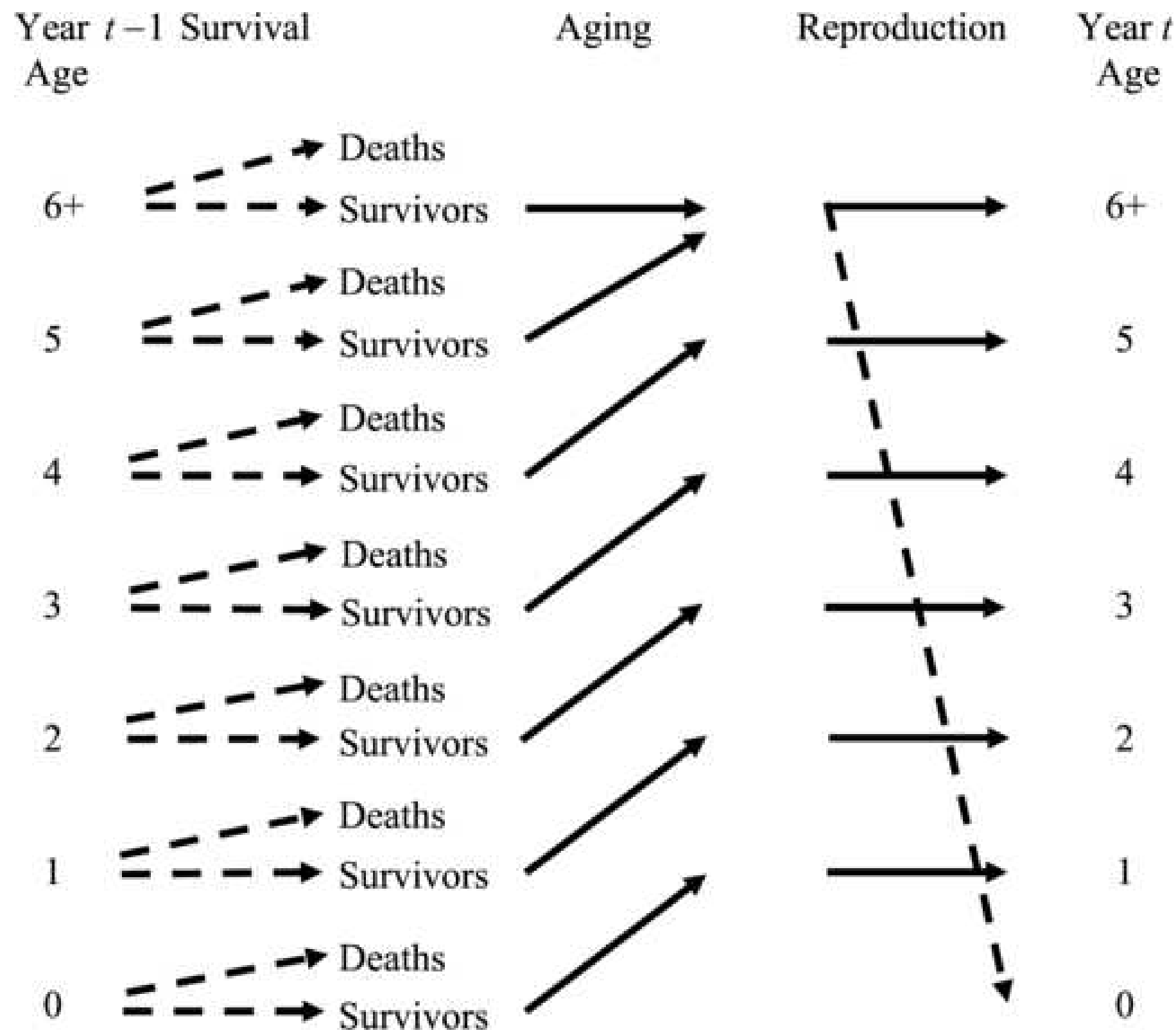

FIG. 1. *Model for British grey seals. For clarity, only one region is shown. Dashed lines indicate stochastic processes, solid lines deterministic processes.*

Hebrides (19 colonies), Outer Hebrides (11 colonies) and the Orkneys (22 colonies). We used this model to investigate whether it was more likely that the differential trends in pup production observed among the four regions were due to density-dependent survival and movement, or to deliberate killing of seals to protect salmon farms.

The model is shown diagrammatically in Figure 1. We can come close to expressing it in the form of a matrix population model by writing $E(\mathbf{n}_t|\mathbf{n}_{t-1}) = \mathbf{B}\mathbf{M}_t\mathbf{A}\mathbf{S}_t\mathbf{n}_{t-1}$, where $\mathbf{n}_t$ and $\mathbf{n}_{t-1}$ are column vectors listing number of individuals in each of seven age categories and four geographical regions. The matrix $\mathbf{B}$ represents reproduction, with the same birth rate for all mature females (age 6 or more); $\mathbf{M}_t$ represents movement of seals aged 5 (those recruiting to the breeding population) between regions, which is modeled as a function of relative densities and distances between sites; $\mathbf{A}$ represents the annual deterministic process of aging; and $\mathbf{S}_t$ is a diagonal matrix of survival probabilities, which are modeled as a function of either pup production or salmon farming activity. Note that the movement matrix $\mathbf{M}_t$ is density dependent, and the survival matrix $\mathbf{S}_t$ is either density dependent or a function of salmon farming activity, so that both vary by year, whereas the aging matrix $\mathbf{A}$ is certainly time independent, and the birth matrix $\mathbf{B}$ is assumed to be. The product $\mathbf{P}_t = \mathbf{B}\mathbf{M}_t\mathbf{A}\mathbf{S}_t$ is the generalized Leslie matrix, a population projection matrix. However, the nonlinearity arising from the density dependence in $\mathbf{M}_t$ invalidates this matrix representation of the model, which needs to be interpreted in the more general framework of (1). The density dependence in $\mathbf{S}_t$ would be allowable in the matrix representation framework, because it is modeled using $\mathbf{n}_{t-1}$, whereas the density dependence in $\mathbf{M}_t$ is modeled using the intermediate state vector $\mathbf{u}_{2,t}$.

Pup production estimates were assumed to be normally distributed, and sequential importance sampling was used to fit the models. Two measures of salmon farming activity were considered: annual salmon production by area, and total staff numbers by area. In the matrix notation, $\mathbf{S}_t$ now becomes $\mathbf{S}(\mathbf{x}_t)$, where $\mathbf{x}_t$ is either salmon production or total staff numbers, or $\mathbf{S}(\mathbf{n}_{t-1})$ for the model in which survival is a function of pup production. Annual salmon production, with a common parameter across regions for its possible effect on seal survival, provided the best fitting model, as judged by Akaike's Information Criterion.

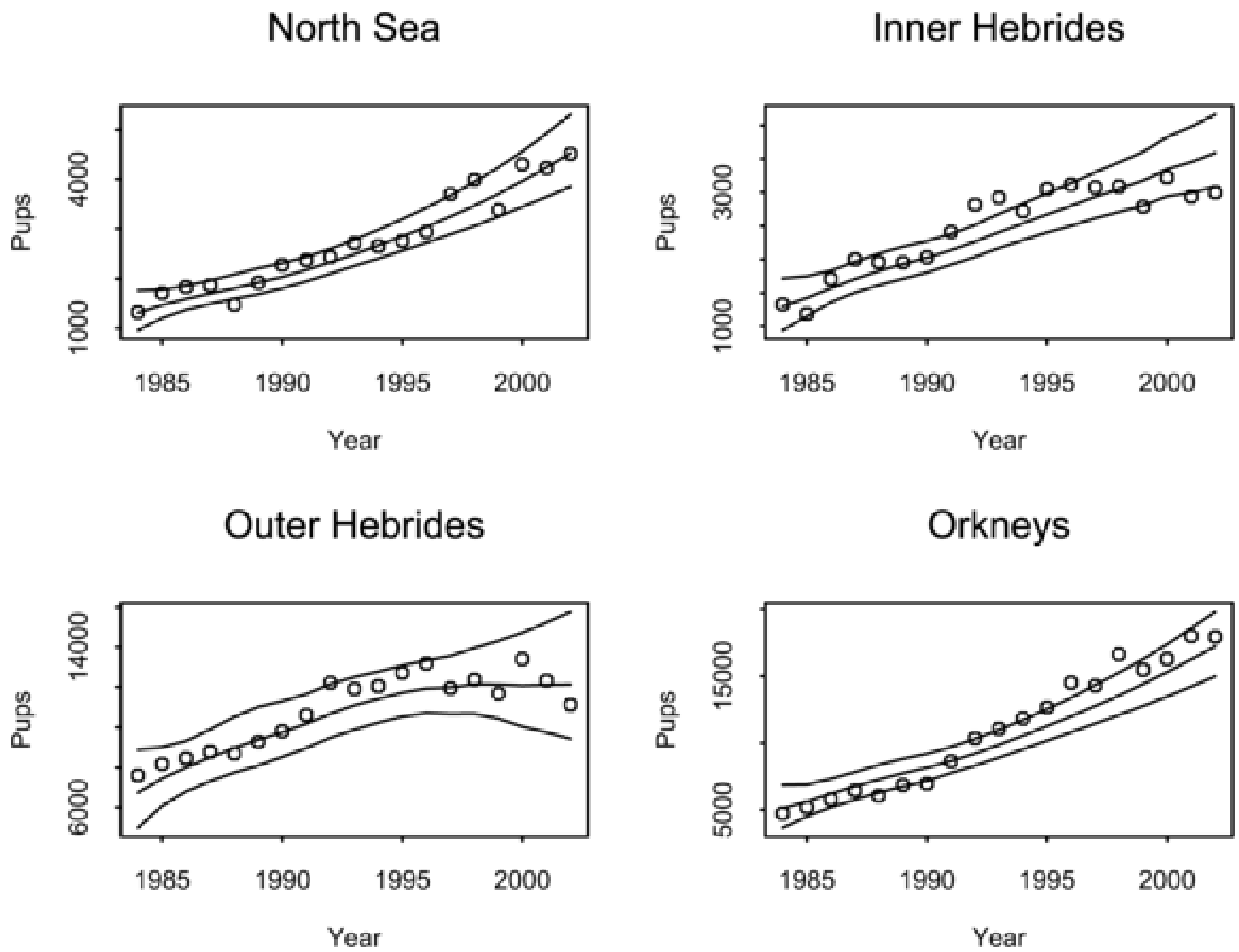

Fig. 2. *Estimates of pup production for model without density dependence and movement, but with annual salmon production as a covariate for modeling survival. Input data are shown by open circles. Mean estimates of pup numbers from the model are shown, together with the* 2.5 *and* 97.5 *percentiles of our estimates.*

Smoothed estimates of numbers of pups in each of the regions are shown in Figure 2 for the model without density dependence and movement, but with annual salmon production as a covariate. This simplistic model is able to fit the time series of pup production estimates reasonably well. However, the estimated levels of deliberate killing are implausibly high, with a best estimate of around 4000 seals shot annually in recent years in the Outer Hebrides. This level of mortality is unlikely to occur undetected, suggesting that the differences in trends in pup production among regions are not due solely to anthropogenic killing.

## 5. DISCUSSION

Ecologists often have a good understanding of the processes regulating the abundance of particular animal or plant populations, but the statistical methods that have traditionally been used to model time series of abundance estimates do not exploit this knowledge. Consequently, the resulting models are ineffective for predicting future changes. They also provide no mechanism for exploring the impact of different management actions, such as harvest strategies or reintroductions of a species into its former range. Explicit models of population processes are needed for this kind of prediction. However, the uncertainty associated with these predictions will not be adequately quantified if these models are not fully embedded into inference. Hence, the risks associated with different management strategies will be unknown. Methods of the type described here are therefore essential for effective management of species and of the ecosystems to which they belong.

In the section on limitations, the issues of continuous-time processes and modeling of communities were raised. These topics offer many research opportunities. Guidelines or rules of thumb for model formulation with associated measures of the extent of nonidentifiability or weak identifiability are needed. For example, given the observation and state vectors, which components of the state vector and which parameters are identifiable? Quoting Harvey (1989, page 205): "The question of identifiability is a fundamental one in statistical modeling. It is particularly important in the context of unobserved components modeling since it is very easy to set up models which are not identifiable." Methods exist for linear regression models (e.g., variance inflation factors and condition numbers of the correlation matrix for predictors; Myers, 1990), for mark-recapture models (Catchpole and Morgan, 1997; Catchpole, Morgan and Freeman, 1998) and for time-invariant structural models (Harvey, 1989), but similar methods have not yet been developed for nonlinear, nonstationary state-space models.

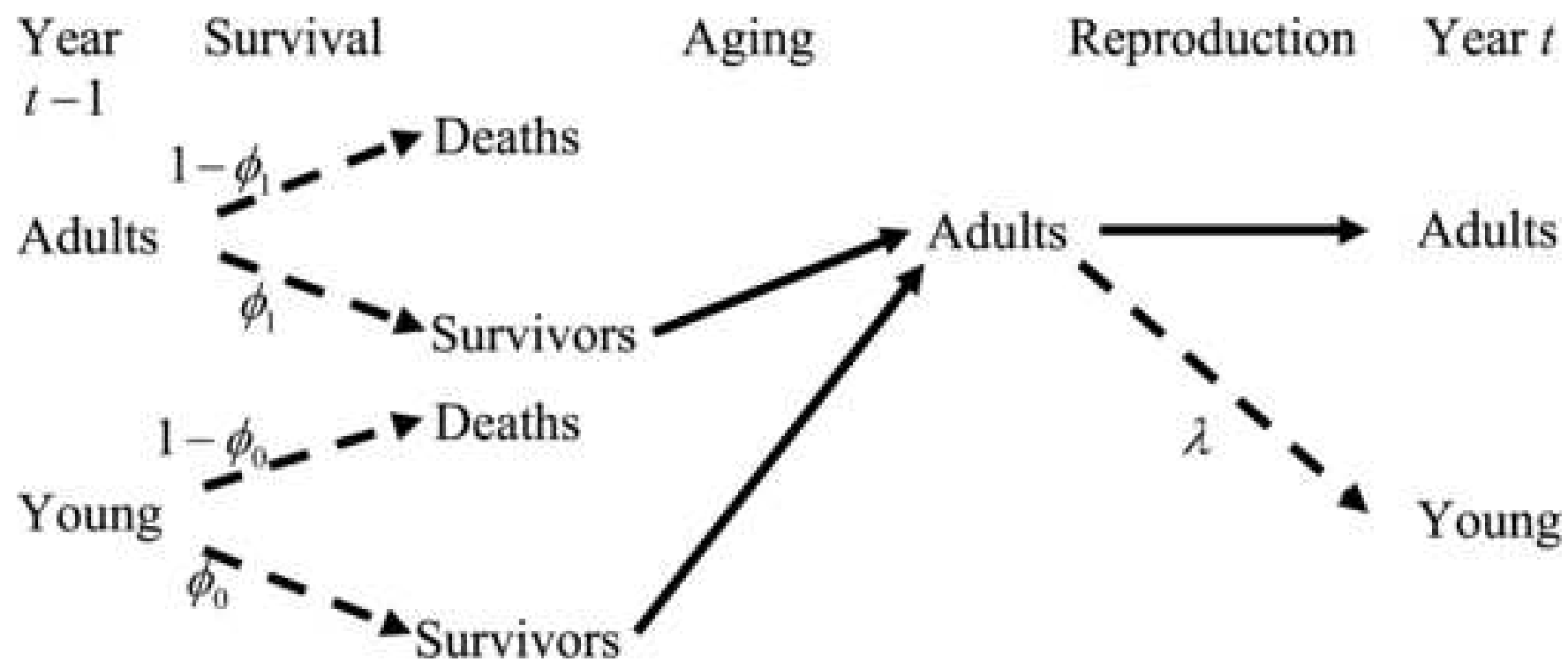

Fig. 3. *Model 1.*

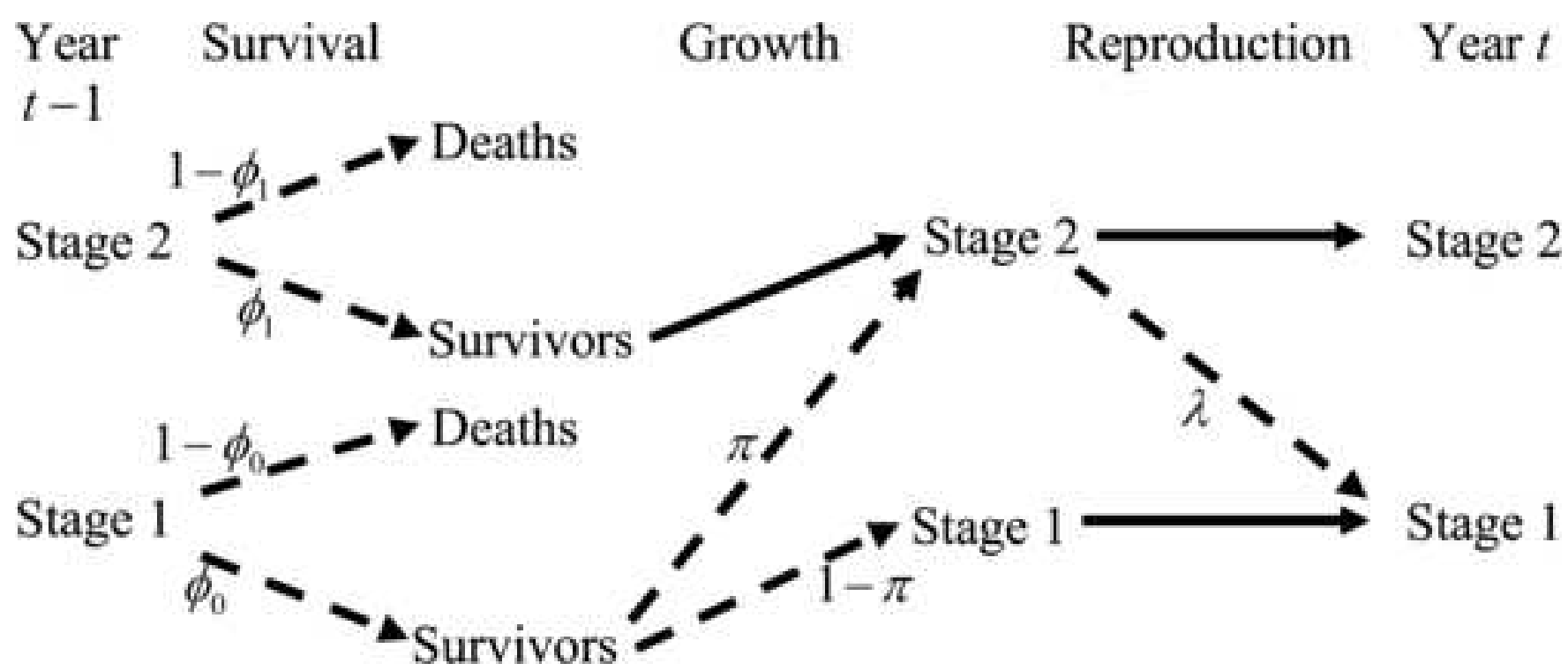

Fig. 4. *Model 2.*

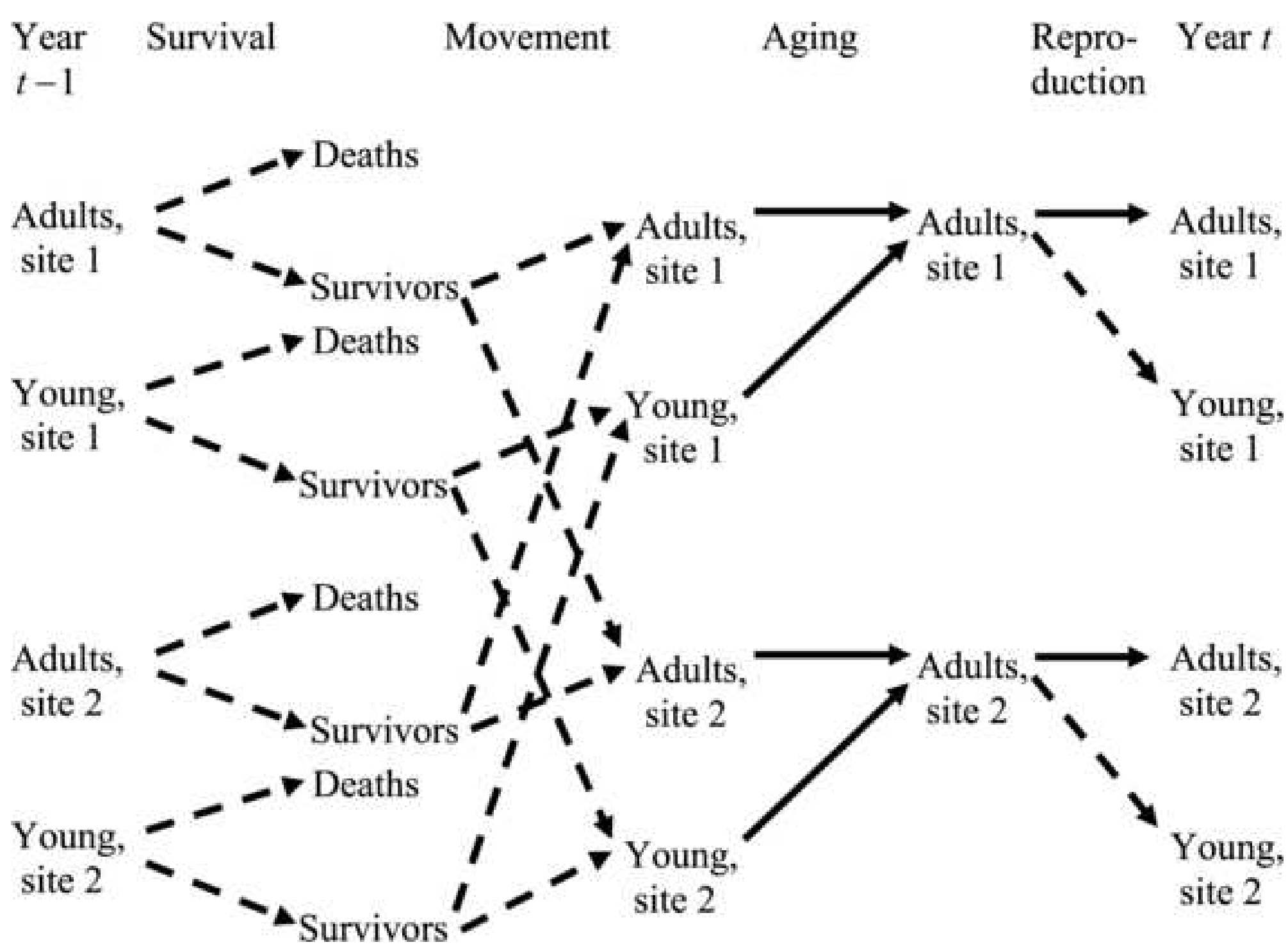

Fig. 5. *Model 3.*

Methods for fitting nonlinear, non-Gaussian state-space models are the subject of much research (e.g., Doucet, de Freitas and Gordon, 2001). Choices include MCMC, SIS and other sequential Monte Carlo procedures. Whereas these are general-purpose computational methodologies, the details of any specific algorithm are bound to be quite complicated. The dimension of the posterior distribution is typically very high, as it includes the static parameters and the state vectors for all years. At the same time, there is often not enough data information to identify all unknowns clearly, leading to strong correlations in the posterior distribution. The combination of high dimensionality with strong correlations is certainly challenging from a computational perspective and there are research opportunities here. In addition, practitioners would benefit from guidance on when to choose one procedure over another, and from the development of software for largely automated fitting of state-space models. WinBUGS (MRC Biostatistics Unit, Cambridge, UK), which uses MCMC, has been successfully used for some complicated state-space models for ecological data (Rivot et al., 2004), but if correlations between parameters and states are high, convergence can be extremely slow (Newman et al., 2007).

Finally, state-space model formulation and inference need to be fully integrated with what Caswell calls a "complete demographic analysis" that includes asymptotic dynamics, transient dynamics and perturbation analysis.

## APPENDIX: DIAGRAMMATIC REPRESENTATIONS OF THE MODELS

Diagrams in Figures 3–5 above are similar to life cycle graphs (Caswell, 2001, pages 56–59). We use dashed arrows to indicate stochastic processes, and solid arrows to indicate deterministic processes. Also shown are the rates associated with the stochastic processes: $\phi_0$ and $\phi_1$ are the survival rates of young and adults, respectively, and $\lambda$ is the mean number of births per adult. For Model 2, $\pi$ is the annual rate of change from stage 1 to stage 2. These rates may themselves be modeled, to allow dependence on environmental variables, population size, numbers of predators, resources, and so on. We can represent Model 1 as Figure 3. Model 2 may be represented by Figure 4. And finally Model 3. Rates have been omitted from Figure 5 to aid clarity.

## ACKNOWLEDGMENTS

We thank the referees and an Editor, whose reviews led to a much improved paper. Stephen T. Buckland also acknowledges the support of the Leverhulme Trust.

## REFERENCES

Berliner, L. M. (1996). Hierarchical Bayesian time series models. In *Maximum Entropy and Bayesian Methods* (K. Hanson and R. Silver, eds.) 15–22. Kluwer, Dordrecht. MR1446713

Besbeas, P., Freeman, S. N. and Morgan, B. J. T. (2005). The potential of integrated population modelling. *Aust. N. Z. J. Stat.* **47** 35–48. MR2144486

Besbeas, P., Freeman, S. N., Morgan, B. J. T. and Catchpole, E. A. (2002). Integrating mark-recapture-recovery and census data to estimate animal abundance and demographic parameters. *Biometrics* **58** 540–547. MR1933532

Besbeas, P., Lebreton, J.-D. and Morgan, B. J. T. (2003). The efficient integration of abundance and demographic data. *Appl. Statist.* **52** 95–102. MR1963214

Buckland, S. T., Magurran, A. E., Green, R. E. and Fewster, R. M. (2005). Monitoring change in biodiversity through composite indices. *Philos. Trans. R. Soc. Lond. Ser. B* **360** 243–254.

Buckland, S. T., Newman, K. B., Thomas, L. and Koesters, N. B. (2004). State-space models for the dynamics of wild animal populations. *Ecological Modelling* **171** 157–175.

Burnham, K. P. and Anderson, D. R. (2002). *Model Selection and Multimodel Inference*: *A Practical Information-Theoretic Approach*, 2nd ed. Springer, New York. MR1919620

Calder, C., Lavine, M., Müller, P. and Clark, J. S. (2003). Incorporating multiple sources of stochasticity into dynamic population models. *Ecology* **84** 1395–1402.

Caswell, H. (2001). *Matrix Population Models*: *Construction*, *Analysis and Interpretation*, 2nd ed. Sinauer Associates, Sunderland, MA.

Catchpole, E. A. and Morgan, B. J. T. (1997). Detecting parameter redundancy. *Biometrika* **84** 187–196.

Catchpole, E. A., Morgan, B. J. T. and Freeman, S. N. (1998). Estimation in parameter-redundant models. *Biometrika* **85** 462–468. MR1649125

Clark, J. S. and Bjørnstad, O. N. (2004). Population time series: Process variability, observation errors, missing values, lags and hidden states. *Ecology* **85** 3140–3150.

Clark, J. S., Ferraz, G. A., Oguge, N., Hays, H. and DiCostanzo, J. (2005). Hierarchical Bayes for structured, variable populations: From recapture data to life-history prediction. *Ecology* **86** 2232–2244.

Collie, J. S. and Sissenwine, M. P. (1983). Estimating population size from relative abundance data measured with error. *Canadian J. Fisheries and Aquatic Sciences* **40** 1871–1879.

Cunningham, C. L., Reid, D. G., McAllister, M. K., Kirkwood, G. P. and Darby, C. D. (2007). Modelling the migration of multiple stocks: a Bayesian state-space model for north-east Atlantic mackerel. *African J. Marine Science*. To appear.

Dempster, A. P., Laird, N. M. and Rubin, D. B. (1977). Maximum likelihood estimation from incomplete data via the EM algorithm (with discussion). *J. Roy. Statist. Soc. Ser. B* **39** 1–38. MR0501537

Doucet, A., de Freitas, N. and Gordon, N., eds. (2001). *Sequential Monte Carlo Methods in Practice.* Springer, New York. MR1847783

Dupuis, J. A. (1995). A Bayesian estimation of movement and survival probabilities from capture–recapture data. *Biometrika* **82** 761–772. MR1380813

Gilks, W. R., Richardson, S. and Spiegelhalter, D. J., eds. (1996). *Markov Chain Monte Carlo in Practice.* Chapman and Hall, London. MR1397966

Gilks, W. R. and Roberts, G. O. (1996). Strategies for improving MCMC. In *Markov Chain Monte Carlo in Practice* (W. R. Gilks, S. Richardson and D. J. Spiegelhalter, eds.) 89–114. Chapman and Hall, London.

Gordon, N. J., Salmond, D. J. and Smith, A. F. M. (1993). Novel approach to nonlinear/non-Gaussian Bayesian state estimation. *IEE Proceedings-F* **140** 107–113.

Green, P. J. (1995). Reversible jump Markov chain Monte Carlo computation and Bayesian model determination. *Biometrika* **82** 711–732. MR1380810

Gudmundsson, G. (1987). Time series models of fishing mortality rates. Report RH-02-87, Raunvisindastofnun Haskolans, Univ. Iceland.

Gudmundsson, G. (1994). Time series analysis of catch-at-age observations. *Appl. Statist.* **43** 117–126.

Gurney, W. S. C. and Nisbet, R. M. (1998). *Ecological Dynamics.* Oxford Univ. Press, New York.

Harvey, A. (1989). *Forecasting, Structural Time Series Models and the Kalman Filter.* Cambridge Univ. Press.

Harwood, J. and Stokes, K. (2003). Coping with uncertainty in ecological advice: Lessons from fisheries. *Trends in Ecology and Evolution* **18** 617–622.

Hilborn, R., Pikitch, E. K. and McAllister, M. K. (1994). A Bayesian estimation and decision analysis for an age-structured model using biomass survey data. *Fisheries Research* **19** 17–30.

Hilborn, R. and Walters, C. J. (1992). *Quantitative Fisheries Stock Assessment: Choice, Dynamics and Uncertainty.* Chapman and Hall, New York.

Hooten, M. B., Wikle, C. K., Dorazio, R. M. and Royle, J. A. (2007). Hierarchical spatio-temporal matrix models for characterizing invasions. *Biometrics* **63** 558–567.

Johnson, D. S. and Hoeting, J. A. (2003). Autoregressive models for capture–recapture data: A Bayesian approach. *Biometrics* **59** 341–350. MR1987401

Kalman, R. E. (1960). A new approach to linear filtering and prediction problems. *Transactions of ASME-J. Basic Engineering* **82** 35–45.

King, R. and Brooks, S. P. (2002a). Model selection for integrated recovery/recapture data. *Biometrics* **58** 841–851. MR1939400

King, R. and Brooks, S. P. (2002b). Bayesian model discrimination for multiple strata capture–recapture data. *Biometrika* **89** 785–806. MR1946510

Lavine, M., Beckage, B. and Clark, J. S. (2002). Statistical modeling of seedling mortality. *J. Agric. Biol. Environ. Stat.* **7** 21–41.

Lebreton, J. D. (1973). Introduction aux modèles mathématiques de la dynamique des populations. *Informatique et Biosphère* 77–116.

Lebreton, J. D. and Isenmann, P. (1976). Dynamique de la population camarguaise de mouettes rieuses *Larus ridibundus* L.: un modèle mathématique. *La Terre et la vie* **30** 529–549.

Lefkovitch, L. P. (1965). The study of population growth in organisms grouped by stages. *Biometrics* **21** 1–18.

Lele, S. R. (2006). Sampling variability and estimates of density dependence: A composite likelihood approach. *Ecology* **87** 189–202.

Leslie, P. H. (1945). On the use of matrices in certain population mathematics. *Biometrika* **33** 183–212. MR0015760

Leslie, P. H. (1948). Some further notes on the use of matrices in population mathematics. *Biometrika* **35** 213–245. MR0027991

Liu, J. S. (2001). *Monte Carlo Strategies in Scientific Computing.* Springer, New York. MR1842342

Liu, J. S. and Chen, R. (1998). Sequential Monte Carlo methods for dynamic systems. *J. Amer. Statist. Assoc.* **93** 1032–1044. MR1649198

McAllister, M. K. and Ianelli, J. N. (1997). Bayesian stock assessment using catch-age data and the sampling-importance sampling algorithm. *Canadian J. Fisheries and Aquatic Sciences* **54** 284–300.

McAllister, M. K., Pikitch, E. K., Punt, A. E. and Hilborn, R. (1994). A Bayesian approach to stock assessment and harvest decisions using the sampling/importance resampling algorithm. *Canadian J. Fisheries and Aquatic Sciences* **51** 2673–2687.

McConnell, B. J., Fedak, M. A., Lovell, P. and Hammond, P. S. (1999). Movements and foraging areas of grey seals in the North Sea. *J. Applied Ecology* **36** 573–590.

Mendelssohn, R. (1988). Some problems in estimating population sizes from catch-at-age data. *Fishery Bulletin* **86** 617–630.

Meyer, R. and Millar, R. B. (1999). Bayesian stock assessment using a state-space implementation of the delay difference model. *Canadian J. Fisheries and Aquatic Sciences* **56** 37–52.

MILLAR, R. B. and MEYER, R. (2000). Non-linear state-space modelling of fisheries biomass dynamics by using Metropolis–Hastings within-Gibbs sampling. *Appl. Statist.* **49** 327–342. MR1824544

MYERS, R. H. (1990). *Classical and Modern Regression with Applications*, 2nd ed. PWS-Kent, Boston.

NEWMAN, K. B. (1998). State-space modeling of animal movement and mortality with application to salmon. *Biometrics* **54** 1290–1314.

NEWMAN, K. B. (2000). Hierarchic modeling of salmon harvest and migration. *J. Agric. Biol. Environ. Stat.* **5** 430–455. MR1812085

NEWMAN, K. B., BUCKLAND, S. T., LINDLEY, S. T., THOMAS, L. and FERNÁNDEZ, C. (2006). Hidden process models for animal population dynamics. *Ecological Applications* **16** 74–86.

NEWMAN, K. B., FERNÁNDEZ, C., BUCKLAND, S. T. and THOMAS, L. (2007). Monte Carlo inference for state-space models of wild animal populations. To appear.

PITT, M. K. and SHEPHARD, N. (1999). Filtering via simulation: Auxiliary particle filters. *J. Amer. Statist. Assoc.* **94** 590–599. MR1702328

POOLE, D. and RAFTERY, A. E. (2000). Inference for deterministic simulation models: The Bayesian melding approach. *J. Amer. Statist. Assoc.* **95** 1244–1255. MR1804247

QUINN, T. J. II and DERISO, R. B. (1999). *Quantitative Fish Dynamics*. Oxford Univ. Press, New York.

RAFTERY, A. E., GIVENS, G. H. and ZEH, J. E. (1995). Inference from a deterministic population dynamics model for bowhead whales (with discussion). *J. Amer. Statist. Assoc.* **90** 402–430.

RIVOT, E., PRÉVOST, E., PARENT, E. and BAGLINIÈRE, J. L. (2004). A Bayesian state-space modelling framework for fitting a salmon stage-structured population dynamic model to multiple time series of field data. *Ecological Modelling* **179** 463–485.

RUBIN, D. B. (1988). Using the SIR algorithm to simulate posterior distributions. In *Bayesian Statistics 3* (J. M. Bernardo, M. H. DeGroot, D. V. Lindley and A. F. M. Smith, eds.) 395–402. Clarendon Press, Oxford.

SCHNUTE, J. T. (1994). A general framework for developing sequential fisheries models. *Canadian J. Fisheries and Aquatic Sciences* **51** 1676–1688.

SULLIVAN, P. (1992). A Kalman filter approach to catch-at-length analysis. *Biometrics* **48** 237–257.

THOMAS, L., BUCKLAND, S. T., NEWMAN, K. B. and HARWOOD, J. (2005). A unified framework for modelling wildlife population dynamics. *Aust. N. Z. J. Stat.* **47** 19–34. MR2134470

THOMPSON, D., HAMMOND, P. S., NICHOLAS, K. S. and FEDAK, M. A. (1991). Movements, diving and foraging behaviour of grey seals (*Halichoerus grypus*). *J. Zoology* **224** 223–232.

TRENKEL, V. M., ELSTON, D. A. and BUCKLAND, S. T. (2000). Calibrating population dynamics models to count and cull data using sequential importance sampling. *J. Amer. Statist. Assoc.* **95** 363–374.

TULJAPURKAR, S. (1997). Stochastic matrix models. In *Structured-Population Models in Marine*, *Terrestrial and Freshwater Systems* (S. Tuljapurkar and H. Caswell, eds.) 59–87. Chapman and Hall, New York.

WALTERS, C. (2002). *Adaptive Management of Renewable Resources.* Blackburn, Caldwell, NJ.

WEST, M. (1993a). Approximating posterior distributions by mixtures. *J. Roy. Statist. Soc. Ser. B* **55** 409–422. MR1224405

WEST, M. (1993b). Mixture models, Monte Carlo, Bayesian updating and dynamic models. In *Computing Science and Statistics*: *Proc. 24th Symposium on the Interface* 325–333. Interface Foundation of North America, Fairfax Station, VA.

WEST, M. and HARRISON, J. (1997). *Bayesian Forecasting and Dynamic Models*, 2nd ed. Springer, New York. MR1482232

WIKLE, C. K. (2003). Hierarchical Bayesian models for predicting the spread of ecological processes. *Ecology* **84** 1382–1394.

WIKLE, C. K., BERLINER, L. M. and CRESSIE, N. (1998). Hierarchical Bayesian space–time models. *Environmental and Ecological Statistics* **5** 117–154.

WOLPERT, R. L. (1995). Comment on "Inference from a deterministic population dynamics model for bowhead whales," by A. E. Raftery, G. H. Givens and J. E. Zeh. *J. Amer. Statist. Assoc.* **90** 426–427.